\documentclass[preprint2]{aastex}
\newcommand \cmsq           {\hbox{cm$^{-2}$}}

\newcommand \kms          {\rm{\hbox{km s$^{-1}$}}}
\newcommand \lam          {$\lambda$}
\newcommand \Lya          {\hbox{Ly$\alpha$}}

\newcommand \mum           {\hbox{$\mu$m}}
\newcommand \pcc           {\hbox{cm$^{-3}$}}

\newcommand \zaz          {{$z_a\kern -1.5pt \approx\kern -1.5pt z_e$}}
\newcommand \zllz         {{$z_a\kern -3pt \ll\kern -3pt z_e$}}
\newcommand \Zsun          {\hbox{Z$_{\odot}$}}
\begin{document}
\title
{\Large\bf Metallicities and Abundance Ratios from \\
Quasar Broad Emission Lines}
\smallskip
\author{Fred Hamann\altaffilmark{1}, K.T.
Korista\altaffilmark{2}, G.J. Ferland\altaffilmark{3},
Craig Warner\altaffilmark{1}, Jack Baldwin\altaffilmark{4}}
\altaffiltext{1}{Department of Astronomy, University of Florida,
211 Bryant Space Science Center, Gainesville, FL 32611-2055,
Internet: hamann@astro.ufl.edu}

\altaffiltext{2}{Department of Physics, University of Western Michigan,
1120 Everett Tower, Kalamazoo, MI 49008-5252}

\altaffiltext{3}{Department of Physics and Astronomy, University of
Kentucky, Lexington, KY 40506}

\altaffiltext{4}{Department of Physics and Astronomy, Michigan State
University, East Lansing, MI 48824-1116}

\begin{abstract}
\normalsize

The broad emission lines (BELs) of quasars and active galactic
nuclei (AGNs) are important diagnostics of the relative abundances
and overall metallicity in the gas. Here we present new
theoretical predictions for several UV BELs. We focus specifically
on the relative nitrogen abundance as a metallicity indicator,
based on the expected secondary enrichment of nitrogen at
metallicities $Z\ga 0.2$ \Zsun . Among the lines we consider,
\ion{N}{3}] $\lambda$1750/\ion{O}{3}] $\lambda$1664, \ion{N}{5}
$\lambda$1240/(\ion{C}{4} $\lambda$1549 + \ion{O}{6}
$\lambda$1034) and \ion{N}{5}/\ion{He}{2} $\lambda$1640 are the
most robust diagnostics. We argue, in particular, that the average
\ion{N}{5} BEL is not dominated by scattered \Lya\ photons from a
broad absorption line wind. We then compare our calculated line
ratios with observations from the literature. The results support
earlier claims that the gas-phase metallicities near quasars are
typically near or several times above the solar value. We conclude
that quasar activity is preceded by, or coeval with, an episode of
rapid and extensive star formation in the surrounding galactic (or
proto-galactic) nuclei. Chemical evolution models of
these environments suggest that, to reach $Z\ga$ \Zsun\ in well-mixed
interstellar gas, the star formation must have begun $\ga$10$^8$~yr
before the observed quasar activity.

\end{abstract}

\newpage

\section{Introduction}

Measurements of the elemental abundances near quasars or, more
generally, active galactic nuclei (AGNs) are important for several
reasons. For example, the metal abundances affect the opacities,
kinematics and overall physical structure of
the various emitting and absorbing regions (e.g., Komossa \& Mathur 2001,
Ferland et al. 1996, hereafter F96 ). On much larger scales,
metal-rich outflows from quasars or their host galaxies might
be important sources of metal enrichment to the inter-galactic gas
at early cosmological times (Friaca \& Terlevich 1998,
Collin \& Zahn 1999a, Sadat, Guiderdoni \&
Silk 2001). Quantifying these effects requires explicit abundance
measurements, ideally at a range of distances from the
active nucleus.

Quasar abundance data also provide information on the chemical
history of the gas and therefore, indirectly, on the epoch and
extent of nearby star formation (see Hamann \& Ferland 1999,
hereafter HF99, for a general review). The metals near
high-redshift quasars might come from some of the first stars
forming in galactic or proto-galactic nuclei after the Big Bang.
The constraints on star formation and chemical evolution {\it
near} quasars are therefore a valuable complement to other high
redshift
studies that use different diagnostics or probe larger
galactic structures (e.g., Lu et al. 1996, Pettini et al. 1999,
Steidel et al. 1999, Prochaska, Gawiser \& Wolfe 2001
and references therein).

In this paper, we develop further the theoretical basis for
deriving abundances from the rest-frame UV broad emission lines
(BELs). We note that although these features form within
$\sim$1 pc of the central AGNs (Kaspi et al. 2000), the derived
abundances are diagnostic of the chemical evolution in
whatever (larger) region processed the gas.

Previous BEL studies have shown that we cannot simply compare a
strong metal line\footnote{Throughout this paper we use the
notation C IV and C III], for example, to identify
the lines C IV~\lam 1549 and C III] \lam 1909.
These designations are unique for
the lines of interest here (summarized in Table 1 below).
Note that the wavelengths listed in Table 1 are often
averages over several transitions in the same multiplet.},
such as \ion{C}{4}~\lam 1549, to \Lya\ to
derive the overall metallicity, $Z$
(e.g., Hamann \& Ferland 1993a, HF99). In numerical
simulations of photoionized BEL regions (BELRs),
the \ion{C}{4}/\Lya\ flux ratio changes by factors of just
a few (and in a non-monotonic way) as the metallicity (C/H)
increases by orders of magnitude. This insensitivity to $Z$ is caused
mainly by the important role of the collisionally-excited
\ion{C}{4} line in cooling the gas; increasing C/H ratios, which might
otherwise increase \ion{C}{4}/\Lya , lead instead to lower gas
temperatures and a nearly constant \ion{C}{4}/\Lya\ flux.
Another major drawback is that \ion{C}{4} and \Lya\ can form in
spatially distinct regions
(Korista et al. 1997a) and, in any given region, their relative
fluxes depend sharply on both the degree of ionization and the shape
of the
ionizing spectrum (see also
Davidson \& Netzer 1979,
Ferland 1999).

The best BEL abundance diagnostics involve ions that are
not important in the cooling and lines that have similar
excitation/emission requirements (forming as much as
possible in the same gas).

\subsection{Nitrogen and Galactic Chemical Evolution}

Nitrogen lines are especially valuable
because the N/O abundance is known to scale with O/H (i.e., $Z$)
in galactic \ion{H}{2} regions (Shields 1976,
Pagel \& Edmunds 1981, Villa-Costas \& Edmunds 1993, van Zee,
Salzer \& Haynes 1998, Izotov \& Thuan 1999). This abundance behavior
is attributed to ``secondary'' N production, whereby N is
synthesized from existing C and O via CNO burning
in stars (Tinsley 1980). The net result is
N/H~$\propto$~(O/H)$^2$~$\propto Z^2$ and
N/O~$\propto$~O/H~$\propto Z$ whenever secondary dominates primary
N production. The \ion{H}{2} region data show that
these scaling relations hold approximately for $Z\ga 0.2$ \Zsun .
The relative nitrogen abundance, e.g. N/O,
can therefore be diagnostic of the overall metallicity even
when direct measures of $Z$ (such as O/H) are not available.

Galactic chemical evolution models predict that the
N production should be delayed relative to O, and to
a lesser extent C, in environments with high star formation
rates and short overall enrichment times, $\la$2 Gyr
(HF99, Hamann \& Ferland 1993b,
hereafter HF93b). The delay is significant if the lifetimes of the
N-producing stars are comparable to the time scale for major increases
in $Z$ (e.g., in O/H, see also Henry, Edmunds \& K\"oppen 2000).
The net result is that N/H and N/O can be below
their solar values at $Z$ = \Zsun . This offset is only weakly
present in the \ion{H}{2} region data (van Zee et al. 1998) because
those data represent more slowly evolving environments. However, the
offset could be substantial in quasars if their host environments
are galactic nuclei, or dense proto-galactic condensations, that
are undergoing rapid evolution (HF99).

The gas-phase N abundance is thus given generally
by\footnote{We prefer to use oxygen
(or some equivalent ``$\alpha$-capture'' element like Ne, Mg or Si)
in these scaling relationships because it is a good tracer
of the overall metallicity driven by massive-star supernovae.
The behavior of N/C, in particular,
might be more complex, depending on the shape of the initial mass
function and the chemical evolution time scale. The complication
is that C/O can also increase (by factors up to 2--3) as the
system evolves, because of the late-time release of C from
intermediate-mass stars (Wheeler, Sneden \& Truran 1989,
HF93b). Secondary N production will respond to this delayed
C enrichment, but the departures of N/H or N/O from Equations 1 \& 2
should be small because the total C+O abundance is in all cases
dominated by O. (We assume implicitly
that secondary N comes from both C and O, and that the primary
yields of these elements are not dependent on $Z$, c.f. Henry
et al. 2000.)},
\begin{equation}
\left[{{\rm N}\over{\rm H}}\right]\ \approx\
2\left[{{\rm O}\over{\rm H}}\right] - q\ \approx\
2\log\left({{Z}\over{{\rm Z}_{\odot}}}\right) - q
\end{equation}
\begin{equation}
\left[{{\rm N}\over{\rm O}}\right]\ \approx\
\left[{{\rm O}\over{\rm H}}\right] - q\ \approx\
\log\left({{Z}\over{{\rm Z}_{\odot}}}\right) - q
\end{equation}
where the square brackets indicate logarithmic ratios relative
to solar, [a/b] = $\log (a/b) - \log (a/b)_{\odot}$, and $q$ is the
logarithmic offset. Values of $q$ range from $\sim$0.0--0.1 for ``slow''
chemical evolution (as reflected in the \ion{H}{2} region data)
to $\sim$0.2--0.5 during the rapid evolution that may occur in massive
galactic nuclei (see the enrichment model in HF99).

\subsection{Quasar Abundance Diagnostics}

Shields (1976) proposed that several ratios of UV intercombination
(semi-forbidden) BELs can provide accurate CNO abundance ratios.
The lines include, for example, \ion{C}{3}] \lam 1909,
\ion{N}{3}] \lam 1750 and \ion{O}{3}] \lam 1664.
Early results gave evidence for an overabundance of N
relative to C and O, indicating $Z\ga$ \Zsun\ by the argument in
\S1.1 (see also Davidson 1977, Osmer 1980, Gaskell, Wampler \& Shields
1981, Uomoto 1984). However, those studies did not consider collisional
deexcitation of the upper energy levels (Baldwin \& Netzer 1978).
The critical densities
(where the rate of collisional deexcitation matches
the rate of radiative decays) are as low as $\sim$$3\times 10^{9}$ \pcc\
for \ion{C}{3}] (see Table 1 below). The observed strength
of \ion{C}{3}] relative to \ion{C}{4} led to the general belief
that BELR densities are $\la$$3\times 10^{9}$~\pcc\
(Davidson \& Netzer 1979). However, more recent line variability
studies (Peterson 1993, Goad, O'Brien \& Gondhalekar
1993, Ferland et al. 1992, Reichert et al. 1994,
Korista et al. 1995, Peterson \& Wandel 1999, Kaspi \& Netzer 1999)
indicate that the BELR is spatially stratified with, quite likely,
a wide range of densities. Different lines form in different
regions depending on their ionization and/or density requirements
(see also Baldwin et al. 1995, Korista \& Goad 2000).
Ferland et al. (1992) used variability results for NGC 5548 to estimate
a density of $\sim$$10^{11}$~\pcc\ in the \ion{C}{4} region,
well above the previous upper limit
inferred from \ion{C}{3}]/\ion{C}{4}.

Partly to overcome this uncertainty, but also to invoke stronger
and thus easier-to-measure BELs, Hamann \& Ferland (1992), HF93b
and F96 developed \ion{N}{5} \lam 1240/\ion{C}{4} and
\ion{N}{5}/\ion{He}{2} \lam 1640 as abundance diagnostics. Their
results indicate that enhanced N abundances and solar
or higher metallicities are typical of BELRs (see also Osmer et al.
1994, Laor et al. 1994 and 1995, Korista et al. 1998, Dietrich et
al. 1999,  Dietrich \& Wilhelm-Erkens 2000, Constantin et al.
2001, Warner et al. 2001).

Independent confirmation of the BEL abundances has come from
quasar associated absorption lines (AALs), which also reveal
typically high metallicities ($Z\ga$ \Zsun ) and probably enhanced
N/C ratios in gas near the quasars (Wampler et al. 1993,
Petitjean, Rauch \& Carswell 1994, Hamann 1997, Petitjean \&
Srianand 1999, HF99). The overall implication is that quasar
environments are metal-rich and thus chemically ``mature'' ---
having already undergone substantial star formation {\it prior} to
the observed quasar epochs. This result appears to hold generally,
even for quasars at the highest measured redshifts (HF99,
Constantin et al. 2001, Warner et al. 2001).

Here we test and further quantify the abundance sensitivity of
various BEL ratios. Throughout this paper, we define solar abundances
by the meteoritic results in Grevessse \& Anders (1989). We
employ atomic data from Verner, Verner \& Ferland
(1996) and from the compilation given in HAZY ---
the documentation supporting the spectral synthesis
code, CLOUDY, which we also use below for line flux calculations
(Ferland et al. 1998). CLOUDY and HAZY
are both freely available on the World Wide Web
(http://www.pa.uky.edu/$\sim$gary/cloudy).

\section{Analytic Estimates of the Line Flux Ratios}

Analytic estimates of
the line flux ratios can yield both physical insight and accurate
abundance results. The theoretical ratio of two collisionally excited
lines emitted from the same volume by two-level atoms is,
\begin{equation}
{{F_1}\over{F_2}} \ = \ Q \ {{n_{l1}}\over{n_{l2}}} \
{{\Upsilon_1\, \lambda_2\, g_{l2}}\over{\Upsilon_2\, \lambda_1\, g_{l1}}}
\ \exp\left({{{\Delta E_2-\Delta E_1}\over{kT_e}}}\right)
\end{equation}
where $Q$ is defined by
\begin{equation}
Q \ \equiv \ {{1+{{n_e}\over{n_{cr2}\, \beta_2}}}\over
{1+{{n_e}\over{n_{cr1}\, \beta_1}}}}
\end{equation}
Stimulated emission and interactions with the continuum radiation
field are
assumed to be negligible. For each line designated 1 or 2,
$F$ is the measured flux, $\lambda$ is the rest wavelength,
$\Delta E$ is the photon energy, $\Upsilon$ is the
energy-averaged collision strength,
$n_l$ and $g_l$ are the number density and statistical weight of
the lower energy state, $\beta$ is the photon escape probability
($0\leq\beta\leq 1$), and $n_{cr}$ is the critical density of the
upper energy state. $T_e$ is the electron temperature and
$n_e$ is the electron density.
Table 1 lists critical densities for the
lines of interest here.

The factor $Q$ in Equation 3 accounts for possible photon
trapping and collisional deexcitation. If the densities are low,
such that $n_e\ll n_{cr}\beta$ for both lines in a given ratio,
then $Q\approx 1$ and collisional deexcitation is not important. If
the densities are high such that $n_e\gg n_{cr}\beta$ for both
lines, then collisional deexcitation {\it is} important and
$Q\approx n_{cr1}\beta_1/n_{cr2}\beta_2$. (The level populations
approach LTE in this limit.) If the line
photons escape freely (e.g., for forbidden or
semi-forbidden lines with low oscillator strengths),
then $\beta_1,\beta_2\approx 1$ and
the correction factor at high densities
is simply $Q\approx n_{cr1}/n_{cr2}$.

The lines become ``thermalized'' and thus lose their abundance
sensitivities\footnote{We can see that the abundance sensitivity is
lost in the
high density, optically thick limit because $Q\rightarrow
n_{cr1}\beta_1/n_{cr2}\beta_2\propto
\Upsilon_2\tau_2/\Upsilon_1\tau_1\propto \Upsilon_2 n_{l2}/\Upsilon_1 n_{l1}$,
where $\tau$ is the line optical depth and $\beta \sim 1/\tau$
for $\tau\gg 1$ (Frisch 1984). Plugging into Equation 3
shows that the line ratios do not depend on the relative abundance
in this regime.} only in the limit where $\beta\ll 1$
{\it and} $n_e\gg n_{cr}\beta$, that is, where the lines are optically
thick {\it and} the density exceeds the ``effective'' critical
density of $n_{cr}\beta$. The abundance sensitivity remains
in all other circumstances, namely, (1) in all cases where the
lines are optically thin -- including the high density limit
(where $n_e\gg n_{cr}\beta$), and (2) at large line optical depths
{\it if} the density is low enough to satisfy
$n_e\ll n_{cr}\beta$ (see also HF99).

If the ions are primarily in their ground states (or ground
multiplets, such that $n_l$ approximates the total ionic
density), then we can rewrite Equation 3 as,
\begin{equation}
{{F_1}\over{F_2}} \ = \
B \ Q \ {{f(X_1^i)}\over{f(X_2^j)}} \
\left({{X_1}\over{X_2}}\right) \
\exp\left({{C}\over{T_4}}\right)
\end{equation}
where $T_4$ is the gas temperature in units of 10$^4$~K,
$f(X_1^i)$ is the fraction of element $X_1$ in ion stage $X_1^i$,
etc., and $X_1/X_2$ is the abundance ratio by number. $B$ and $C$
contain the various rate coefficients and physical constants from
Equation 3. Table 2 provides values of $B$ and $C$ for the line
ratios of interest here. $B$ has a weak temperature
dependence (leftover from the ratio of collision strengths) that is
negligible for our applications (c.f.. Netzer 1997). The tabulated
$B$ values assume $T_e = 10^4$~K. Table 2
also lists the flux
ratios derived from Equation 5 in the
two limiting optically
thin cases, $Q=1$ and $Q= n_{cr1}/n_{cr2}$, assuming
$f(X_1^i)/f(X_2^j)=1$ and solar abundances. The temperatures used
for these flux ratios are based on the numerical simulations in \S3.1
below, namely, $T_4=0.8$ for \ion{N}{2}]/\ion{C}{2}],
$T_4=2.2$ for the ratios involving \ion{N}{5}, and $T_4=1.5$
for all others. Note that the $Q= n_{cr1}/n_{cr2}$ flux ratios listed
in Table 2 for the permitted lines are not expected to occur in BELRs,
because these lines are very optically thick (\S3.3.3, Figure 6 in
Hamann et al. 1995).

\section{Numerical Simulations}

Modern numerical simulations yield more reliable abundance results
because they self-consistently treat the ionization,
temperature, radiative transfer and line emission
in realistic BELR clouds.
We use the numerical code CLOUDY, version 96.00 (Ferland et al. 1998)
to predict BEL fluxes in different circumstances. We assume the
emitting clouds are free of dust and photoionized by the
quasar continuum radiation. The degree of ionization depends
mainly on the shape of the incident
spectrum and the ionization parameter,
$U\equiv\Phi_{\rm H}/cn_{\rm H}$, where $n_{\rm H}$ is the volume
density of hydrogen particles (\ion{H}{1}+\ion{H}{2}) and
$\Phi_{\rm H}$ (units = \cmsq\ s$^{-1}$) is the flux of
hydrogen-ionizing photons incident on the clouds.
The total column density in all of
our calculations is $\log N_{\rm H}({\rm cm}^{-2}) =
23.5$ (c.f. Ferland \& Persson 1989).

\subsection{A Typical Cloud Structure}

Figure 1 illustrates the temperature and ionization structure
of a ``typical'' BELR cloud having $U=0.1$,
$n_{\rm H} = 10^{10}$ \pcc , solar abundances,
and an incident spectrum defined by Mathews \& Ferland (1987,
hereafter MF87). We consider the effects of other continuum shapes
and cloud properties below. The model cloud in Figure 1 produces
strong emission
in most of the lines of interest here (see \S3.2
below). The main
exceptions are \ion{C}{2}] and \ion{C}{3}], for
which $n_{\rm H} = 10^{10}$ \pcc\ is above their critical densities
(Table 1). Pairs of ions that overlap
substantially in Figure 1 are good candidates for abundance
diagnostics. The figure shows, for example, that N$^{+4}$
resides within the He$^{++}$ zone, that the C$^{+3}$ and O$^{+5}$
zones bracket the N$^{+4}$ region, and that C$^{+2}$--N$^{+2}$
and C$^{+3}$--N$^{+3}$ define cospatial zones that both overlap
with O$^{+2}$ (see also Netzer 1997).

Figure 2 shows more specifically where the different lines form
within our ``typical'' BELR cloud (as defined above for Figure 1).
In particular, the plot shows the line emissivities, $J$, multiplied
by
the local line escape probabilities, $\beta$, as a function of
spatial depth into the cloud.
The emissivities are further multiplied by the spatial depth, $D$,
in Figure 2 to offset the tendency for the logarithmic scale to exaggerate
the emission from small depths. Lines whose emission regions
significantly overlap in Figure 2 may be good abundance
diagnostics (see also F96).

\subsection{Predicted Line Strengths and Ratios}

Figures 3 and 4 show theoretical BEL equivalent widths and flux ratios,
respectively, emitted by clouds with different $n_{\rm H}$
and $\Phi_{\rm H}$. Other input parameters are the same as Figure 1.
(See Korista et al. 1997a, Baldwin et al. 1995 and Ferland
1999 for many more plots similar to Figure 3.)

We next consider the effects of different
continuum shapes and metallicities ranging from $Z = 0.2$ \Zsun\
to 10 \Zsun . Table 3 lists the abundances used for each $Z$
(see also the Appendix).
The metals are all scaled keeping solar proportions,
except for nitrogen, which is scaled according to Equation 1 with
$q=0$. Note that by choosing $q=0$ we will derive
maximum nitrogen line strengths for a given
$Z$, leading to conservatively low estimates of $Z$ when
compared to the measured line ratios (\S7 below).
The change in relative helium abundance is small and therefore
unimportant. We scale He from solar such that the
change in He mass fraction, $\Delta Y$, equals the change in
metallicity, i.e., $\Delta Y/\Delta Z = 1$ (Baldwin et al. 1991,
HF93b and references therein).

If BELRs include simultaneously wide ranges in
$n_{\rm H}$ and $\Phi_{\rm H}$, then each line will
form primarily in regions that most favor its emission.
This situation has been
dubbed the Locally Optimally-emitting Cloud (LOC) model (Baldwin
et al. 1995). It is consistent with observational results
and a natural extension of
multi-zone models (Rees, Netzer \& Ferland 1989, Peterson 1993,
Brotherton et al. 1994a, Baldwin et al. 1996, Hamann et al. 1998).
It also has the tremendous advantage of
not requiring specific knowledge of $n_{\rm H}$ and $\Phi_{\rm H}$.
The total line emission is simply an integral over the
$n_{\rm H}$--$\Phi_{\rm H}$ plane (e.g., Figure 3), with weighting
factors that specify the relative numbers of clouds with high versus
low $n_{\rm H}$ and high versus low $\Phi_{\rm H}$ (see also
Ferguson et al. 1997). We adopt equal weighting per decade
in the $n_{\rm H}$--$\Phi_{\rm H}$ distribution, which is already known
to provide a good match to typical AGN spectra (Baldwin 1997,
Korista \& Goad 2000). The full parameter ranges included in the LOC
integration are $7\leq \log n_{\rm H}({\rm cm}^{-3})\leq 14$ and
$17\leq \log \Phi_{\rm H}({\rm cm}^{-2}{\rm s}^{-1})\leq 24$.

Figure 5 shows predicted line flux ratios as a function of $Z$
for three different incident spectra in the LOC
calculations\footnote{The results in Figure 5 are not
sensitive to the LOC integration limits. For example, revising the
limits to $8\leq \log n_{\rm H}({\rm cm}^{-3})\leq 12$ and
$18\leq \log \Phi_{\rm H}({\rm cm}^{-2}{\rm s}^{-1})\leq 24$
changes the line ratios by $<$0.1 dex at a given $Z$.}.
The incident continua are i) MF87, as in Figures
1--4, ii) a ``hard'' power law
with index $\alpha = -1.0$ ($f_{\nu} \propto\nu^{\alpha}$)
across the infrared through X-rays, and iii) a
segmented power law that approximates recent observations,
namely, $\alpha = -0.9$ from 50 keV to 1 keV ,
$\alpha = -1.6$ from 1 keV to 912 \AA , and
$\alpha = -0.6$ from 912 \AA\ to 1 \mum\ (Zheng et al. 1996,
Laor et al. 1997). All of the model spectra have a steep decline
at wavelengths longer than 1--10 microns to prevent
significant free-free heating in dense clouds (Ferland et al. 1992).
The true spectral shape is not well
known. The MF87 spectrum remains a good guess for the continuum
{\it incident on BELRs}, in spite of its differences with recent
observations (Korista, Ferland \& Baldwin 1997b). In any event,
the three spectra used for Figure 5 span a wide range
of possibilities ---
from a strong ``big blue bump'' in the MF87 continuum to
none at all in the $\alpha = -1.0$ power law.
Figure 5 shows that the uncertainty in the continuum shape
has little effect on the line ratios considered here (except,
perhaps, for \ion{N}{5}/\ion{He}{2}, which is discussed
extensively by F96 and HF99, see also \S3.3.4 and \S4 below).

It is important to note that the results in Figure 5 do
not depend on the validity of the LOC model. Most of the line
ratios, notably \ion{N}{3}]/\ion{O}{3}], do not vary sharply with
$n_{\rm H}$ or $\Phi_{\rm H}$ in the regions where the
lines individually are strong (Figures 3 and 4). The LOC integrations
are therefore simply a convenient way of summarizing the
information in plots like Figure 4. For line ratios that do
have significant $n_{\rm H}$--$\Phi_{\rm H}$ sensitivities,
such as \ion{N}{5}/\ion{C}{4}, the LOC results should minimally
represent the average properties of diverse quasar samples.

\subsection{Parameter Sensitivities}

\subsubsection{Density and Ionizing Flux}

Most of the lines are weak near the
lower right and upper left extremes in Figure 3 because the
ionization parameter is too low or too high, respectively.
Some lines are also weak in
the upper right because, for any given $U$, their emission can
be suppressed at high $n_{\rm H}$ (\S2, Korista et al. 1997a).

Figure 4 shows the sensitivity of the line {\it ratios} to
$n_{\rm H}$ and $\Phi_{\rm H}$. We discuss the
implications for abundance estimates in \S4 below.

\subsubsection{Turbulence and Continuum Pumping}

One poorly known parameter in BELR clouds is the internal
Doppler velocity, $v_D$. Our calculations assume
$v_D$ is strictly thermal (i.e., $v_D\approx v_{th}=\sqrt{2kT/m}$).
Larger Doppler velocities would lead to lower line optical depths
and reduced thermalization effects at high densities
(\S2 and \S3.3.3). Larger $v_D$ can also enhance the line emission via
``continuum fluorescence,'' that is, resonant absorption of continuum
photons followed by radiative decays. If the absorption lines are
optically thick, then this fluorescent contribution  to the BELs will
depend much more on $v_D$ than the abundances (e.g., Ferguson, Ferland
\& Pradhan 1995, Hamann et al. 1998).

We examine these effects by varying $v_D$ from
0 to 3000 \kms\ in the fiducial cloud model portrayed in Figure 1.
Bottorff et al. (2000) present more extensive calculations using
different cloud parameters. The overall result is that continuum
pumping
is insignificant in \ion{C}{4} and all of the intercombination
lines for any $v_D \leq 3000$ \kms .
(In \ion{C}{4} the collisional
contribution
is overwhelming; in the intercombination lines the
optical
depths are too low at large $v_D$.)
Pumping is similarly not important for the other
lines considered here as long as $v_D \la 300$ \kms . Pumping begins
to exceed collisional excitation\footnote{Note that continuum pumping
can be relatively more
important for a given $v_D$ in environments with lower $n_{\rm H}$
and $N_{\rm H}$, such as the narrow emission line regions of AGNs
(Ferguson et al. 1995, Netzer 1997).} in \ion{N}{3},
\ion{O}{3} and \ion{O}{6} only for $v_D\ga 300$ to 800 \kms , and in
\ion{N}{5} and \ion{C}{3} for $v_D\ga 1000$ \kms .
The relevant
line ratios are even less sensitive to $\Delta v$ than the
line strengths individually. For example,
\ion{N}{5}/(\ion{C}{4}+\ion{O}{6}) changes by $<$0.1 dex across the
range $0\leq\Delta v\leq 1000$ \kms .
Since the actual value of $v_D$ in BELRs is unknown (Bottorff et al.
2000), we will assume that $v_D$ is not large enough
to affect the line ratios of interest here.

\subsubsection{Line Optical Depths}

Our calculations indicate that the intercombination lines
are all optically thin, and the permitted lines are all optically
thick, in the regions where their emission is strongest
(see also Ferland et al. 1992, Hamann et al. 1998). The
surprising lack of density dependence in
the \ion{N}{4}]/\ion{C}{4} ratio in Figure 5
(c.f.. Table 2) is caused by the fact that $\beta n_{cr}$ for \ion{N}{4}]
is similar to $\beta n_{cr}$ for \ion{C}{4}, so the lines behave
similarly with density. This behavior
depends on \ion{C}{4} becoming thermalized at high densities, where
it no longer has a direct sensitivity to the abundances (\S2).
Thermalization can
occur in other lines as well, most notably \ion{C}{3}, but also in some
of the intercombination lines. However, thermalization in
limited parts of the $n_{\rm H}$--$\Phi_{\rm H}$ plane is not a
problem for abundance studies if BELRs have, as expected,
a range of physical conditions (Ferland et al. 1992,
Brotherton et al. 1994a, Baldwin et al. 1995, Baldwin et al. 1996).
Each line's emission should be dominated by clouds
where thermalization is not important.

\subsubsection{Metallicity and Abundance Ratios}

The predicted line flux ratios in Figure 5 scale almost linearly with
the abundance ratios (N/O and N/C) and therefore also with
the metallicity
(because $Z\propto$ N/O $\propto$ N/C by Equation 2).
Overall, there is
remarkably good agreement between the numerical
and analytic results.
For example, the numerical results at
$Z$ = \Zsun\ lie
generally between the analytic ratios derived
for $Q=1$
and $Q= n_{cr1}/n_{cr2}$ in Table 2.

It is interesting to note that \ion{N}{5}/\ion{He}{2}
also scales roughly linearly with $Z$ in Figure 5, even though
N/He $\propto$ $Z^2$ (\S1). This behavior is tied to
the energy balance in BELR clouds; higher
metallicities increase the efficiency of metal-line cooling and
thus lower the temperature. The temperature drops by exactly
the amount needed to
maintain the total line emission and preserve the overall energy
balance. \ion{N}{5}/\ion{He}{2} reacts strongly to
temperature changes because
it compares a collisionally excited
line to a recombination
line (c.f. Equations 2, 3 and 6 in HF99).
The rise in \ion{N}{5}/\ion{He}{2} with
increasing N/He is therefore moderated by the dropping temperature.
The net result is that \ion{N}{5}/\ion{He}{2} scales roughly like
N/O $\propto$ $Z$ (i.e., like the enhancement
of N relative to the other metal coolants; see also
F96, Ferland et al. 1998 and HF99).

\section{Preferred Line Ratio Diagnostics}

The theoretical accuracy of any line ratio abundance diagnostic
depends on a variety of factors, such as
1) the temperature sensitivity of the ratio
(Equation 5), 2) the similarity
of the ionization potentials and critical densities,
3) the extent to which the line-emitting regions overlap
spatially
(Figures 1--3), and
4) the constancy of the ratio across a range of plausible $n_H$
and $\Phi_H$ (Figure 4).

Based on these criteria, the most reliable abundance diagnostic
among the intercombination lines is \ion{N}{3}]/\ion{O}{3}]. Note
that O$^{+2}$ can occupy a larger region than N$^{+2}$
(Figure 1), and therefore low column density clouds might contribute
significant \ion{O}{3}] emission but little \ion{N}{3}]. Comparing
observed values of \ion{N}{3}]/\ion{O}{3}] to our calculations
should therefore yield, in the worst case, lower limits to N/O
and thus $Z$.

\ion{N}{3}]/\ion{C}{3}] is also useful, but it may be less robust 
than
\ion{N}{3}]/\ion{O}{3}] because
it has a stronger temperature dependence and the
relevant critical densities differ by a factor of nearly 2 (see
Tables 1 and 2).
The net result is a greater sensitivity to
non-abundance parameters (e.g. Figure 4). For example,
a larger/smaller proportion of low density clouds in the
LOC
integration would have led to $\sim$2 times lower/higher predicted
\ion{N}{3}]/\ion{C}{3}] ratios in Figure 5. We also recall (footnote 6)
that N/C may not be as clearly tied to the metallicity as N/O.

Among the collisionally-excited permitted lines, the most robust
diagnostics should be \ion{N}{3}/\ion{C}{3} and
\ion{N}{5}/(\ion{C}{4}+\ion{O}{6}) because of their relatively
weak temperature dependencies. Note, in particular, that the
ionization and excitation energies of C$^{+3}$ and
O$^{+5}$
bracket those of N$^{+4}$ along the lithium isoelectronic
sequence. The combined ratio
\ion{N}{5}/(\ion{C}{4}+\ion{O}{6})
therefore has lower sensitivities
to temperature, ionization and continuum shape compared to the
separate ratios \ion{N}{5}/\ion{C}{4} and \ion{N}{5}/\ion{O}{6}.

\ion{N}{5}/\ion{He}{2} is sensitive to the temperature and ionizing
continuum shape (\S3.3.4) and, like \ion{N}{3}]/\ion{O}{3}], it may
under represent the N/He abundance because N$^{+4}$ can occupy a
smaller region than He$^{++}$ (inside the He$^{++}$ zone, F96, Figure 1).
Nonetheless, HF93b and F96 showed that
by adopting BELR parameters that maximize the predicted
\ion{N}{5}/\ion{He}{2} (for example the $\alpha = -1.0$ power law
in Figure 5), this ratio can yield firm lower limits on N/He
and therefore $Z$.

\section{Is \ion{N}{5} Enhanced by \Lya ?}

Several studies have noted that the \ion{N}{5} emission line
might be enhanced by \Lya\ photons scattered in a broad
absorption line (BAL) wind
(Surdej \& Hutsemekers 1987, Turnshek 1988, Hamann, Korista
\& Morris
1993, Turnshek et al. 1996, Krolik \& Voit 1998).
The underlying premise is that all
quasars have BAL outflows, but they are detected
only when the gas happens to lie along our line of sight
to the continuum source.
The BAL gas is believed to be
located outside the radius
of both the continuum
emission region and the BELR (see, for example,
Figure 4
in Cohen et al. 1995). If BAL winds include N$^{+4}$ ions
outflowing at $\sim$5900 \kms\ (corresponding to the wavelength
difference between \Lya\ and \ion{N}{5}), then these ions
can resonantly scatter \Lya\ BEL (and continuum)
photons into our line of sight --
thus increasing the observed \ion{N}{5} emission strength.
This scattering contribution from a BAL wind is not included
in our calculations (\S2 and \S3).

However, there is strong evidence that the scattering contributions
to \ion{N}{5} BELs are typically small.
In particular, the average \ion{N}{5} BAL intercepts only $\sim$30\%
of the incident \Lya\ flux (based the depth of the average \ion{N}{5}
BAL trough extrapolated across \Lya ; Hamann \& Korista 1996).
Therefore, only $\sim$30\% of the \Lya\ photons incident on
BAL regions are available for scattering into the \ion{N}{5} BEL.
(Krolik \& Voit's [1998] estimate of the scattered \ion{N}{5} flux
is at least $\sim$3 times too large because they assumed that
\ion{N}{5} BALs intercept all of the incident \Lya\ photons.)

The actual scattering contributions to {\it measured} BELs depend
further on 1) the scattered line profiles (i.e., how the scattered
line flux is distributed in velocity), and 2) the global
covering factor of the BAL wind (i.e., the fraction of the sky
covered by BAL gas as seen by the central continuum source).
The total scattered flux scales linearly with the covering factor.
Detailed line scattering calculations assuming an
average BAL velocity profile and a nominal 12\% covering factor yield
total scattered \ion{N}{5} fluxes that are only $\sim$25\% of
the average \ion{N}{5} BEL
measured in quasars
(Hamann \& Korista 1996). The 12\% covering factor is based
on the $\sim$12\% detection frequency of BALs in quasar samples
(Foltz et al. 1989 and 1990).
There is some evidence from polarization studies that the average
covering factor might be 20\% to 30\% (Schmidt \& Hines 1999). However,
30\% is probably a firm upper limit because larger covering factors
would lead to too much scattered flux filling in the bottoms of observed
BAL troughs (Hamann et al. 1993).

The calculations by Hamann \& Korista (1996) also show that the
scattered line flux from a BAL wind is likely to appear over
a wider range of velocities than typical BELs. For scattering
to dominate the {\it measured}
\ion{N}{5} BEL, the N$^{+4}$ ions in BAL winds, with velocities
reaching typically 10,000--20,000 \kms , must scatter \Lya\ plus
continuum photons into BEL profiles that have average half-widths of
only $\sim$2500 \kms\ (e.g., Brotherton et al. 1994b).
Isotropic scattering of the \Lya\ photons can therefore be ruled
out, because that would produce flat-topped \ion{N}{5}
emission profiles with half-widths of $\sim$5900 \kms . Scattered
continuum photons would appear over an even wider range of line
velocities ($\pm$20,000 \kms\ based
on the average \ion{N}{5} and \ion{C}{4} BAL troughs, Hamann \&
Korista 1996). Highly anisotropic wind geometries are required, but
that causes other problems if the scattered flux dominates,
namely, BELs that should be
broader in quasars with observed BALs compared to quasars without
BALs (see Hamann \& Korista 1996 and Hamann et al. 1993 for specific
calculations). These differences
are not observed in quasar BELs (Weymann et al. 1991).

Another important point is that
observed \ion{N}{5} emission profiles closely resemble those of
other BELs (e.g., \ion{C}{4}) measured in the same
spectrum (Osmer et al. 1994, Laor et al. 1995,
Constantin et al. 2001).
This occurs naturally for lines forming in more or
less the same BELR gas. However, the profile similarities become
a serious problem if \ion{N}{5} is attributed to scattering in a
high-velocity wind, while the other BELs are collisionally excited
in a separate region (the usual BELR) whose velocity field is
believed to be smaller and mostly
{\it not} radial (T\"urler \& Courvoisier 1997, Korista et al. 1995,
Peterson \& Wandel 1999).

Finally, reverberation studies indicate that high ionization
BELs such as \ion{N}{5} and \ion{He}{2}
form generally closer to the continuum source than \Lya\ and other
lower ionization lines (Korista et al. 1995, Dietrich \& Kollatschny
1995, Wanders et al. 1997, O'Brien et al. 1998).
In contrast, \ion{N}{5} BALs
are observed to suppress the \Lya\ emission flux, presumably because
the \ion{N}{5} BAL region lies {\it outside} the \Lya\ BELR radius.
These results are apparently contradictory if \ion{N}{5} BELs are
attributed to scattering in a BAL wind.

All of these difficulties suggest that measured \ion{N}{5} BELs
are controlled by
normal BELR emission, and {\it not} by scattered
flux from a BAL
outflow.

\section{Observed Line Ratios}

Figure 5 includes observed BEL flux ratios from several
studies (see figure caption). The data represent
quasars over a range
of redshifts and luminosities. For example,
Dietrich \& Wilhelm-Erkens (2000) measured luminous
quasars at redshifts $z\ga 3$, while Laor et al. (1994, 1995) and
Wills et al. (1995) observed relatively less luminous
sources at $z\la 1$. The data plotted from Boyle
(1990) derive from a mean spectrum that emphasizes
relatively low luminosity quasars with $z\sim 1$--2.

We include data points from the literature only if both lines
in a given ratio were measured. This criterion
selects against weak lines and might bias the plotted data
toward, for example, smaller
\ion{N}{5}/\ion{He}{2} ratios or larger \ion{N}{3}]/\ion{C}{3}].
However, inspection of the original references indicates that any
such biasing is not severe. In particular, if we include upper limits
on undetected weak lines, the resulting ratios are still consistent
with the ranges shown for directly measured values. Ratios involving
two weak lines, such as \ion{N}{3}]/\ion{O}{3}], should also be unbiased
because such lines are generally either both present or both absent
in the data. In addition, the measurements from the mean spectrum
in Boyle (1990) are not subject to
these selection biases, yet they are similar to the other
average values in Figure 5. We therefore expect the data in
Figure 5 to be representative of the original quasar samples.

Several of the observed BELs can be blended with adjacent features.
We use ``deblended'' measurements in all cases except for \ion{C}{3}],
which can be blended with
\ion{Si}{3}] \lam 1892 and \ion{Al}{3}] \lam 1859. Efforts to
deblend the \ion{C}{3}] complex (Laor et al. 1995, Steidel \& Sargent
1991) have shown  that \ion{C}{3}] contributes typically 65\% to
90\% to the total flux. If the original study made
no attempt to deblend these lines, we multiplied the flux reported
for the overall blend by 0.8 to approximate the \ion{C}{3}]
contribution alone.

The observed value of \ion{N}{4}]/\ion{O}{4}] in Figure 5 is based on a
single \ion{N}{4}]/\ion{O}{4}] measurement by Laor et al. (1995).
This measurement is unreliable because \ion{N}{4}] is weakly detected
and \ion{O}{4}] is severely blended with \ion{Si}{4}
$\lambda\lambda$1394,1403 emission.

\section{Abundance Results}

The observed line ratios are shown in Figure 5 on the left-hand side
of each panel for convenience. The metallicities implied these data
are determined by noting the value of $Z$ where the observed ratios
intersect the theoretical curves. We conclude from these comparisons
that quasar metallicities are typically near or a few
times above the solar value. In particular,
the most robust diagnostics, such as \ion{N}{3}]/\ion{O}{3}]
and \ion{N}{5}/(\ion{C}{4}+\ion{O}{6}) (\S4), indicate typically
$Z\sim 1$--3 \Zsun . \ion{C}{3}]/\ion{N}{3}] and \ion{C}{3}/\ion{N}{3}
provide similar results. \ion{N}{5}/\ion{He}{2} also strongly supports
$Z \ga$ \Zsun , and is consistent with $Z\sim 1$--3 \Zsun\
if we adopt the relatively ``hard'' power law continuum, $\alpha = -1$.
\ion{N}{4}]/\ion{O}{3}] is less reliable theoretically (\S2 and \S3.1)
and shows considerable scatter among the observations
(see also Warner et al. 2001). We therefore do not consider this
ratio further. We also disregard
the other \ion{N}{4}]
line ratios because they have even larger
uncertainties (\S4 and \S6).

Actual quasar metallicities could be as much as 2--3
times higher than these estimates because the theoretical
calculations (Table 2 and Figures 4 and 5) are based on $q=0$ in
Equation 1. In fact, $q\sim 0.2$--0.5 may be more appropriate for
the rapid chemical enrichment occurring near quasars in galactic
nuclei (HF93b, \S1.1). The results presented here are therefore
broadly consistent with quasar metallicities\footnote{Some previous
BEL studies favored
the high end of this metallicity range (e.g., HF93b, Ferland et
al. 1996, Korista et al. 1998, Dietrich et al. 1999) largely because
they
assumed $q>0$ for the nitrogen enrichment.} being typically in
the range $Z\sim 1$--9 \Zsun .

\section{Discussion}

It is important to keep in mind that the metallicities
discussed here are based on the assumptions of secondary nitrogen
production and N $\propto Z^2$ scaling (\S1.1).
Actual abundance ratios might vary
if, for example, there was a recent short-duration
starburst (which could magnify differences in the time-dependent
yields of C, N and O; Lu, Sargent \& Barlow 1996, Coziol et al.
1999, Henry et al. 2000).
It is also possible that the scaling relations in Equations 1 and 2
break down at $Z\gg$ \Zsun\ if the yields of C and O depend
significantly on $Z$ (Henry et al. 2000). In addition,
rare quasars with very strong nitrogen lines (Osmer 1980,
Baldwin et al. 2001) might not have metallicities as high
as Figure 5 would suggests; their environments might be unusually
nitrogen rich because of localized ``pollution'' by Wolf-Rayet stars,
luminous blue variables or rare planetary nebulae
(Davidson et al. 1986, Kobulnicky et al. 1997, Tajitsu et al. 1999).

Nonetheless, an approximate N $\propto Z^2$ scaling seems well
established for most circumstances (\S1.1), and so the
metallicities inferred from Figure 5 should be approximately correct.
It is also reassuring that the results derived here from the BELs
are in good agreement with metallicity estimates based on
intrinsic absorption lines (\S1.2), which do not in any way rely
on assumptions about the nitrogen production.
The metallicities implied by all these diagnostics are typically
near or a few
times above the solar value.

If these gas-phase metallicities result from the normal evolution
of surrounding stellar populations, then those populations must
already be fairly evolved at the observed quasar epoch (but see
Collin \& Zahn 1999b for an alternative hypothesis). In
particular, chemical evolution models using  ``standard'' initial
mass functions show that $Z\ga\Zsun$ occurs in well-mixed
interstellar gas only after most of the original gas has been
converted into stars and stellar remnants (e.g., HF93b). At high
redshifts, this evolution must have occurred quickly (with
a high star formation rate) to achieve $Z\ga\Zsun$ by the time the
quasars became observable.
For example, at redshift $z\sim 5$ the
available evolution time is
only
$\sim$1--2 Gyr, depending on cosmological models (see Figure 1
in
HF99).

Such rapid evolution may seem extreme by solar neighborhood
standards, but it is well within the parameters derived
independently for dense proto-galactic condensations (Ostriker \&
Gnedin 1997, Cen \& Ostriker 1999) and (at least the cores of)
giant elliptical galaxies (HF93b, Friaca \& Terlevich 1998,
Granato et al. 2001, HF99 and references therein).
It is also
consistent with other studies suggesting that the degree of
chemical enrichment in any cosmic structure depends more on its
density (i.e. the depth of its gravitational potential) than on
its redshift (Pettini 2001). Quasars provide direct
observational
evidence for high metallicities and rapid evolution
in dense galactic or proto-galactic nuclei at high redshifts
(see also Dietrich \& Hamann 2001).
Complementary evidence
has come from observations of quasar host galaxies
(Nolan et al. 2001), which suggest that the
hosts are typically
massive ellipticals that have already formed
by the epoch of peak
quasar activity ($z\sim 2.5$).

Chemical enrichment models designed specifically for dense galactic
nuclei (HF93b, Friaca \& Terlevich 1998) suggest that the
time needed to reach $Z\sim$ \Zsun\ in the gas is only of order
10$^8$
yr. We conclude, therefore, that quasar host galaxies begin an
episode of major star formation at least $\sim$10$^8$ yr before
the onset of visible quasar activity.
These star forming episodes
might be among the first to occur in collapsed structures after
the
Big
Bang. It is clearly desirable now to extend the quasar
abundance
analysis to the highest
possible redshifts, and compare the results across a wide range
of redshifts, luminosities
and other quasar/host galaxy properties.

\section{Summary}

We present new theoretical calculations showing the abundance
sensitivities of various UV broad emission lines in quasar/AGN
spectra. We focus on ratios involving C, N and O, and rely on the
secondary enrichment of nitrogen to provide a crude metallicity
indicator. The most robust abundance probes among the line ratios
we consider are \ion{N}{3}]/\ion{O}{3}],
\ion{N}{5}/(\ion{C}{4}+\ion{O}{6}) and \ion{N}{5}/\ion{He}{2}.
Comparing our calculations to observed data from the literature
suggests that quasar metallicities are typically near or several
times higher than the solar value. These metallicities probably
result from the rapid evolution of stellar populations in
galactic (or proto-galactic) nuclei at high redshifts ($z> 2$--3).
Chemical evolution models of these environments suggest that, to reach
$Z\ga$ \Zsun\ in the gas, the episode of
major star formation began
$\ga$$10^8$ yr
before the quasars become observable.

\acknowledgments
We are grateful to Matthias Dietrich, Hagai Netzer, Bassem Sabra and
an anonymous referee for comments on this manuscript.
FH acknowledges financial support from the NSF via grant AST 99-84040.
GF thanks the NSF for support through AST-0071180 and NASA for support
through its LTSA program (NAG5-8212).

\bigskip\bigskip
\centerline{\bf APPENDIX}

Throughout this paper, we adopt the common usage definition of
metallicity,
\begin{equation}
{{Z}\over{\Zsun}} \ \approx \ {{\rm metals/H}\over
{({\rm metals/H})_{\odot}}} \ \approx \ {{\rm O/H}\over
{({\rm O/H})_{\odot}}}
\end{equation}
where O, H and ``metals'' represent abundances by number.
However, $Z$ is strictly speaking a mass fraction, such
that $X+Y+Z\equiv 1$, where $X$ and $Y$ represent the fractions in
hydrogen and helium. If we define
\begin{equation}
\xi \ \equiv \ {{\rm metals/H}\over
{({\rm metals/H})_{\odot}}}
\end{equation}
to simplify the notation,
and adopt $\Delta Y/\Delta Z = 1$ (\S3.2),
then it is easy to show that
the correct relationship between the
mass fraction, $Z$,
and the scale factor, $\xi$, is
\begin{equation}
{{Z}\over{\Zsun}} \ = \ {{\xi}\over{X_{\odot}+Y_{\odot}+
(2\xi -1)Z_{\odot}}}
\end{equation}
where $X_{\odot}$, $Y_{\odot}$ and $Z_{\odot}$ are the Solar mass
fractions
($X_{\odot}+Y_{\odot}+Z_{\odot} = 1$). This expression yields a
limiting value of $Z/\Zsun \rightarrow 25$ ($Z\rightarrow 0.5$)
for $\xi\rightarrow \infty$. It is actually $\xi$ that we
call the ``metallicity'' and refer to as ``$Z$'' throughout this paper
(everywhere except this Appendix, e.g. in Table 3).
For $\xi \la 3$, the approximation $Z/\Zsun \approx \xi$ \ is accurate
to within 10\%. However, at the highest value of $\xi =10$
in Table 3, the corresponding mass fraction $Z$ is just 7.4 \Zsun .

\newpage
\parskip=0pt
\leftskip=0.2in
\parindent=-0.2in
\noindent{\bf REFERENCES}\bigskip

Baldwin, J.A. 1997, in Emission Lines in Active Galaxies: New Methods
and Techniques, IAU Colloq. 159, Eds. B.M. Peterson, F.-Z. Cheng, A.S.
Wilson, ASP Conf. Ser. 113, 80

Baldwin, J.A., Ferland, G.J., Hamann, F., Korista, K.T., \& Warner,
C. 2001, in preparation

Baldwin, J.A., Ferland, G.J., Korista, K.T., Carswell, R.F., Hamann,
F., Phillips, M.M., Verner, D., Wilkes, B.J., \& Williams, R.E. 1996,
\apj, 461, 664

Baldwin, J.A., Ferland, G.J., Korista, K.T., \& Verner, D. 1995, \apj,
455, 119

Baldwin, J.A., Ferland, G.J., Martin, P.G., Corbin, M.R., Cota, S.A.,
Peterson, B.M., Slettebak, A. 1991, \apj, 374, 580

Baldwin, J.A., \& Netzer, H. 1978, \apj, 226, 1

Bottorff, M., Ferland, G., Baldwin, J., \& Korista, K. 2000, \apj,
542, 644

Boyle, B.J. 1990, \mnras, 243, 231

Brotherton, M.S., Wills, B.J., Francis, P.J., \& Steidel, C.C. 1994a, \apj,
430, 495

Brotherton, M.S., Wills, B.J., Steidel, C.C., \& Sargent, W.L.W. 1994b,
\apj, 423, 131

Cen, R., \& Ostriker, J.P. 1999, \apj, 519, L109

Cohen, M.H., Ogle, P.M., Tran, H.D., Vermeulen, R.C., Miller, J.S.,
Goodrich, R.W., \& Martel, A.R. 1995, \apj, 448, L77

Collin, S., \& Zahn, J.P. 1999a, \apss, 265, 501

Collin, S., \& Zahn, J.P. 1999b, \aap, 344, 433

Constantin, A., Shields, J.C., Hamann, F., Foltz, C.B., \& Chaffee, F.H.
2001, \apj, submitted

Coziol, R., Carlos Reyes, R.E., Consid\`ere, S., Davoust, E., Contini,
T. 1999, 345, 733

Davidson, K. 1977, \apj, 218, 20

Davidson, K.D., Dufour, R.J., Walborn, N.R., \& Gull, T.R. 1986,
\apj, 305, 867

Davidson, K., \& Netzer, H. 1979, Rev. Mod. Phys., 51, 715

Dietrich, M., et al. 1999, \aap, 352, L1

Dietrich, M., \& Hamann, F. 2001, in Astrophysical Ages and Time
Scales, in press (astro-ph/0104180)

Dietrich, M., \& Kollaschny, W. 1995, \aap, 303, 405

Dietrich, M., \& Wilhelm-Erkens, U. 2000, \aap, 354, 17

Ferguson, J.W., Ferland, G.J., \& Pradhan, A.K. 1995, \apj, 438, L55

Ferguson, J.W., Korista, K.T., Baldwin, J., Verner, D. 1997, \apj,
487, 122

Ferland, G.J. 1999, in Quasars and Cosmology, Eds. G.J.
Ferland, J.A. Baldwin, ASP Conf. Ser. 162, 147

Ferland, G.J., Baldwin, J.A., Korista, K.T., Hamann, F., Carswell,
R.F., Phillips, M., Wilkes, B., \& Williams, R.E. 1996, \apj, 461, 683, (F96)

Ferland, G.J., \& Persson, S.E. 1989, \apj, 347, 656

Ferland, G.J., Peterson, B.M., Horne, K., Welsch, W.F., \&
Nahar, S.N. 1992, \apj, 387, 95

Ferland, G.J., Korista, K.T., Verner, D.A., Ferguson, J.W., Kingdon, J.B.,
\& Verner, E.M. 1998, \pasp, 110, 761

Foltz, C.B., Chaffee, F.H., Hewitt, P.C., Weymann, R.J., Anderson, S.A.,
\& MacAlpine, G.M. 1989, \aj, 98, 1959

Foltz, C.B., Chaffee, F.H., Hewitt, P.C., Weymann, R.J., \& Morris,
S.L. 1990, \baas, 22, 806

Friaca, A.C.S., \& Terlevich, R.J. 1998, \mnras, 298, 399

Frisch,  H. 1984, in Methods of Radiative Transfer, ed. W. Kalkofen
(Cambridge: Cambridge Univ. Press), 65

Gaskell, C.M., Wampler, E.J., \& Shields, G.A. 1981, \apj, 249, 443

Gnedin, O.Y., \& Ostriker, J.P. 1997, \apj, 487, 667

Goad, M.R., O'Brien, P.T., \& Gondhalekar, P.M. 1993, \mnras, 263, 149

Granato, G.L., Silva, L., Monaco, P., Panuzzo, P., Salucci, P., De Zotti,
G., \& Danese, L. 2001, \mnras, 324, 757

Grevesse, N, \& Anders, E. 1989, in AIP Conf. Proc. 183, Cosmic Abundances
of Matter, ed. C.I. Waddington (NewYork:AIP), 1

Hamann, F. 1997, \apjs, 109, 279

Hamann, F., Cohen, R.D., Shields, J.C., Burbidge, E.M., Junkkarinen, V.T.,
\& Crenshaw, D.M. 1998, \apj, 496, 761

Hamann, F., \& Ferland, G.J. 1992, \apj, 391, L53

Hamann, F., \& Ferland, G.J. 1993a, Rev. Mex. Astron. Astrofis., 26, 53

Hamann, F., \& Ferland, G.J. 1993b, \apj, 418, 11, (HF93b)

Hamann, F., \& Ferland, G.J. 1999, \araa, 37, 487, (HF99)

Hamann, F., \& Korista, K.T. 1996, \apj, 464, 158

Hamann, F., Korista, K.T., \& Morris, S.L. 1993, \apj, 415, 541

Hamann, F., Shields, J.C., Ferland, G.J., \& Korista, K.T. 1995,
\apj, 454, 61

Henry, R.B.C., Edmunds, M.G., \& K\"oppen, J. 2000, \apj, 541, 660

Izotov, Y.I., \& Thuan, T.X. 1999, \apj, 511, 639

Kaspi, S., \& Netzer, H. 1999, \apj, 524, 71

Kaspi, S., Smith, P.S., Netzer, H., Moaz, D., Jannuzi, B.T., \&
Giveon, U. 2000, \apj, 533, 631

Kobulnicky, H.A., Skillman, E.D., Roy, J.-R., Walsh, J.R., \& Rosa,
M.R. 1997, \apj, 477, 679

Komossa, S., \& Mathur, S. 2001, in press with \aap

Korista, K.T., et al. 1995, \apjs, 97, 285

Korista, K.T., Baldwin, J., \& Ferland, G.J. 1998, \apj, 507, 24

Korista, K.T., Baldwin, J., Ferland, G.J., \& Verner, D. 1997a, \apjs,
108, 401

Korista, K.T., \& Goad, M.R. 2000, \apj, 536, 284

Korista, K.T., Ferland, G.J., \& Baldwin, J.A. 1997b, \apj, 487, 555

Krolik, J.H., \& Voit, G.M. 1998, \apj, 497, 5

Laor, A., Bahcall, J.N., Jannuzi, B.T., Schneider, D.P., Green, R.F.,
\& Hartig, G.F. 1994, \apj, 420, 110

Laor, A., Bahcall, J.N., Jannuzi, B.T., Schneider, D.P., \& Green, R.F.
1995, \apjs, 99, 1

Laor, A., Fiore, F., Elvis, M., Wilkes, B.J., McDowell, J.C. 1997,
\apj, 477, 93

Lu, L., Sargent, W.L.W., Barlow, T.A., Churchill, C.W., \& Vogt,
S.S. 1996, \apjs, 107, 475

Mathews, W.G., \& Ferland, G.J. 1987, \apj, 323, 456

McLeod, K.K., Rieke, G.H., \& Storrie-Lombardi, L.J. 1999, \apj, 511, L67

Netzer, H. 1997, Astrophys. Sp. Sci., 248, 127

Nolan, L.A., Dunlop, J.S., Kukula, M.J., Hughes, D.H., Boroson, T.,
\& Jimenez, R. 2001, \mnras, 323, 385

O'Brien, P., et al. 1998, \apj, 509, 1630

Osmer, P.S. 1980, \apj, 237, 666

Osmer, P.S., Porter, A.C., \& Green, R.F. 1994, \apj, 436, 678

Pagel, B.E.J., Edmunds, M.G. 1981, \araa, 19, 77

Petitjean, P., Rauch, M., \&
Carswell, R.F. 1994, \aap, 291, 29

Petitjean, P., \& Srianand, R. 1999, \aap, 345, 73

Peterson, B.M., 1993, \pasp, 105, 247

Peterson, B.M., \& Wandel, A. 1999, \apj, 521, 95

Pettini, M. 2001, Proc. of the ``Promise of FIRST'' Symp., 12-15,
December, 2000, ed. G.L. Pillbratt, J. Cernicharo, A.M. Heras, T.
Prusti, \& R. Harris, Toledo, Spain

Pettini, M., Ellison, S.L., Steidel, C.C., Bowen, D.V. 1999, \apj,
510, 576

Prochaska, J.X., Gawiser, E., \& Wolfe, A.M. 2001, \apj, 552, 99

Rees, M.J., Netzer, H., \& Ferland, G.J. 1989, \apj, 347, 640

Reichert, G.A., et al. 1994, \apj, 425, 582

Sadat, R., Guiderdoni, B. Silk, J. 2001, \aa, 369, 26

Schmidt, G.D., \& Hines, D.C. 1999, \apj, 512, 125

Shields, G. 1976, \apj, 204, 330

Steidel, C.C., Adelberger, K.L., Giavalisco, M., Dickenson, M., \&
Pettini, M. 1999, \apj, 519, 1

Steidel, C.C., \& Sargent, W.L.W. 1991, \apj, 382, 433

Surdej, J., \& Hutsemekers, D. 1987, \aap, 177, 42

Tajitsu, A., Tamura, S., Yadoumaru, Y., Weinberger, R., K\"oppen, J.
1999, \pasp, 111, 1157

Tinsley, B. 1980, Fund. of Cosmic Phys., 5, 287

T\"urler, M., Courvoisier, T.J.-L. 1998, \aap, 329, 863

Turnshek, D.A. 1988, in QSO Absorption Lines: Probing the Universe,
eds. J.C. Blades, D.A. Turnshek, C.A. Norman, Cambridge, UK:Cambridge
Univ. Press, p. 348

Turnshek, D.A., Kopko, M., Monier, E.M., Noll, D., Espey, B.R., \&
Weymann, R.J. 1996, \apj, 463, 110

Uomoto, A. 1984, \apj, 284, 497

van Zee, L., Salzer, J.J., \& Haynes, M.P. 1998, \apj, 497, L1

Verner, D., Verner, E.M., \& Ferland, G.J. 1996, Atomic Data and Nucl.
Data Tables, 64, 1

Vila-Costas, M.B., \& Edmunds, M.G. 1993, \mnras, 265, 199

Wampler, E.J., Bergeron, J., \& Petitjean, P. 1993, \aap, 273, 15

Wanders, I., et al. 1997, \apjs, 113, 69

Warner, C., Hamann, F., Shields, J.C., Constantin, A., Foltz, C., \&
Chaffee, F.H. 2001, \apj, in press

Weymann, R.J, Morris, S.L., Foltz, C.B., Hewitt, P.C. 1991, \apj,
373, 23

Wheeler, J.C., Sneden, C., \& Truran, J.W. 1989, \araa, 27, 279

Wills, B.J., et al. 1995, \apj, 447, 139

Zheng, W., Kriss, G.A., Telfer, R.C., Grimes, J.P., Davidsen, A.F.
1997, \apj, 475, 469

\vfill\eject
\bigskip\bigskip\bigskip
\noindent{\bf FIGURE CAPTIONS}\bigskip
\parskip=0pt
\leftskip=0pt
\parindent=1.5em

Figure 1. ---  Ionization fractions and gas
temperature, $T_4$, versus spatial depth into a ``typical''
BELR cloud ($U=0.1$, $n_H = 10^{10}$ \pcc , solar abundances
and an incident spectrum defined by MF87). The temperature is
indicated in the upper panel by
the dotted curve and by the vertical scale at the right. Ionization
fractions are plotted in the upper panel for helium ions (solid
curves) and hydrogen (dashed), and in the lower panel for nitrogen
(solid curves), carbon (dashed) and oxygen (dash-dot). See \S3.1
\medskip

Figure 2. --- Depth-weighted line emissivities, multiplied by the
escape probability (D$\beta$J), are plotted against spatial depth
into the same ``typical'' BELR cloud described in Figure 1. The lines
can be identified be reference to Table 1. The solid curves
represent lines of nitrogen, the dashed curves represent carbon
lines, the dash-dot curves represent oxygen lines, and the dotted
curve represents HeII.
\medskip

Figure 3. --- Theoretical rest-frame equivalent widths (all
measured relative to the continuum at 1216 \AA ) as a
function $n_{\rm H}$ and $\Phi_{\rm H}$ in BELR clouds. All other
parameters are the same as Figure 1. Contours appear every 0.25 dex,
with the lowest (bold) contour corresponding to 0 dex and other solid
contours marking the 1 or 2 dex boundaries. The ionization parameter,
$U$, is constant along diagonals from the lower left to the upper
right in each panel.
\medskip

Figure 4. --- Theoretical BEL flux ratios for clouds with
different $n_{\rm H}$ and $\Phi_{\rm H}$, based on the same calculations
as Figure 3. The bold outer contours define the boundary within which
both lines in the ratio have rest-frame equivalent widths $\geq$1 \AA .
Other contours are defined as in Figure 3.
\medskip

Figure 5. --- Theoretical line flux ratios derived
from LOC integrations are shown as a function of the metallicity,
$Z$/\Zsun , for three different incident continuum shapes.
The solid curve in each panel corresponds to
the MF87 spectrum, while the dashed and dotted curves
represent the $\alpha = -1.0$ and segmented power laws, respectively
(see \S3.2). The symbols in each panel mark the average measured
line ratios from different literature sources: $\ast$ = Wills et al.
(1995), $\Diamond$ = Laor et al. (1994 and 1995),
$\Box$ = Dietrich \& Wilhelm-Erkens (2000),
$\times$ = Boyle (1990), and $\triangle$ = Uomoto (1984). The
symbols are shown at arbitrary horizontal locations in each plot.
The vertical bars associated with each symbol show the range of ratios
in the middle $\sim$2/3 of the measured distribution (after excluding
the highest $\sim$1/6 and lowest $\sim$1/6 of measured values).

\vfill\clearpage

\begin{table}
\begin{center}
\caption{Critical Densities$^a$}
\begin{tabular}{lc}
\noalign{\vskip 4pt}
\tableline\tableline
\noalign{\vskip 2pt}
Transition& $n_{cr}$(\pcc )\cr
\noalign{\vskip 2pt}
\tableline
\noalign{\vskip 4pt}
C~II] \lam 2326&  3.16e09\cr
C~III \lam 977&  1.59e16\cr
C~III] \lam 1909&  1.03e10\cr
C~IV \lam 1549&  2.06e15\cr
\noalign{\vskip 4pt}
N~II] \lam 2142&  9.57e09\cr
N~III \lam 991&  8.09e15\cr
N~III] \lam 1750&  1.92e10\cr
N~IV] \lam 1486&  5.07e10\cr
N~V \lam 1240&  3.47e15\cr
\noalign{\vskip 4pt}
O~III \lam 834&  2.14e16\cr
O~III] \lam 1664&  3.13e10\cr
O~IV \lam 789&  1.17e16\cr
O~IV] \lam 1401&  1.12e11\cr
O~VI \lam 1034&  5.53e15\cr
\noalign{\vskip 4pt}
Si~III] \lam 1892&  3.12e11\cr
\noalign {\vskip 2pt}
\tableline
\noalign {\vskip 4pt}
\multispan{2}{$^a$For $T_e = 10^4$ K.\hfill}
\end{tabular}
\end{center}
\end{table}

\begin{table}
\begin{center}
\caption{Analytic Line Flux Ratios}
\begin{tabular}{lccccc}
\tableline\tableline
\noalign{\vskip 2pt}
& & & & \multispan{2} \hfil ----- ${F_1}/{F_2}^a$ ----- \hfil\cr
\noalign{\vskip 1pt}
Ratio& $B$& $C$& & $~Q=1$& $Q={{n_{cr1}}\over{n_{cr2}}}$\cr
\noalign{\vskip 2pt}
\tableline
\noalign{\vskip 4pt}
N~II]/C~II]&  0.33& $-$0.53& & 0.04\rlap{5}& 0.14\cr
\noalign{\vskip 12pt}
N~III]/C~III]& 0.36& $-$0.69& & 0.06\rlap{0}& 0.11\cr
\noalign{\vskip 1pt}
N~III]/O~III]& 2.19& +0.42& & 0.36& 0.22\cr
\noalign{\vskip 1pt}
N~III]/Si~III]& 0.06\rlap{7}& $-$0.62& & 0.12& 0.00\rlap{7}\cr
\noalign{\vskip 8pt}
C~III]/Si~III]& 0.19& +0.07& & 1.98& 0.06\rlap{5}\cr
\noalign{\vskip 8pt}
N~IV]/O~III]& 8.82& +1.04& & 0.56& 0.90\cr
\noalign{\vskip 1pt}
N~IV]/O~IV]& 4.58& +0.59& & 0.85& 0.39\cr
\noalign{\vskip 1pt}
N~IV]/C~IV& 0.28& $-$0.39& & 0.05\rlap{6}& 1.4e$-$6\cr
\noalign{\vskip 8pt}
N~III/C~III&  0.30& +0.21& & 0.11& 0.05\rlap{5}\cr
\noalign{\vskip 1pt}
N~III/O~III&  1.80& +2.73& & 1.40& 0.53\cr
\noalign{\vskip 8pt}
N~V/C~IV& 0.95& $-$2.32& & 0.08\rlap{7}& 0.15\cr
\noalign{\vskip 1pt}
N~V/O~VI& 1.09& +2.32& & 0.39& 0.25\cr
\noalign{\vskip 1pt}
N~V/(C~VI+O~VI)& \nodata& \nodata& & 0.07\rlap{1}& 0.09\rlap{4}\cr
\noalign {\vskip 3pt}
\tableline
\noalign {\vskip 4pt}
\multispan{6}{$^a$The flux ratios assume
$f(X_1^i)/f(X_2^j)=1$ and solar \hfill}\cr
\multispan{6}{\ \ \ \ abundances, plus $T_4=0.8$ for
\ion{N}{2}]/\ion{C}{2}], $T_4=2.2$ for\hfill}\cr
\multispan{6}{\ \ \ \ the \ion{N}{5} ratios, and $T_4=1.5$ for
all others.\hfill}\cr
\end{tabular}
\end{center}
\end{table}

\begin{table}
\begin{center}
\caption{Model Abundances$^a$}
\begin{tabular}{cccc}
\noalign{\vskip 4pt}
\tableline\tableline
\noalign{\vskip 2pt}
$Z^b$& N/O& ~~N~~~& He\cr
\noalign{\vskip 2pt}
\tableline
\noalign{\vskip 4pt}
 0.2&  0.2&    0.0\rlap{4}&    0.94\cr
 0.5&  0.5&    0.2\rlap{5}&    0.96\cr
 1.0&  1.0&    1.0&     1.00\cr
 2.0&  2.0&    4.0&     1.07\cr
 5.0&  5.0&  \llap{2}5.0&     1.29\cr
\llap{1}0.0&  \llap{1}0.0&  \llap{10}0.0&    1.66\cr
\noalign {\vskip 2pt}
\tableline
\noalign {\vskip 4pt}
\multispan{4}{$^a$All quantities are rela-\hfill}\cr
\multispan{4}{\ \ \ tive to solar values.\hfill}\cr
\multispan{4}{$^b$$Z$ is the metals scale\hfill}\cr
\multispan{4}{\ \ \ factor, $\xi$ (see Appendix).\hfill}\cr
\end{tabular}
\end{center}
\end{table}

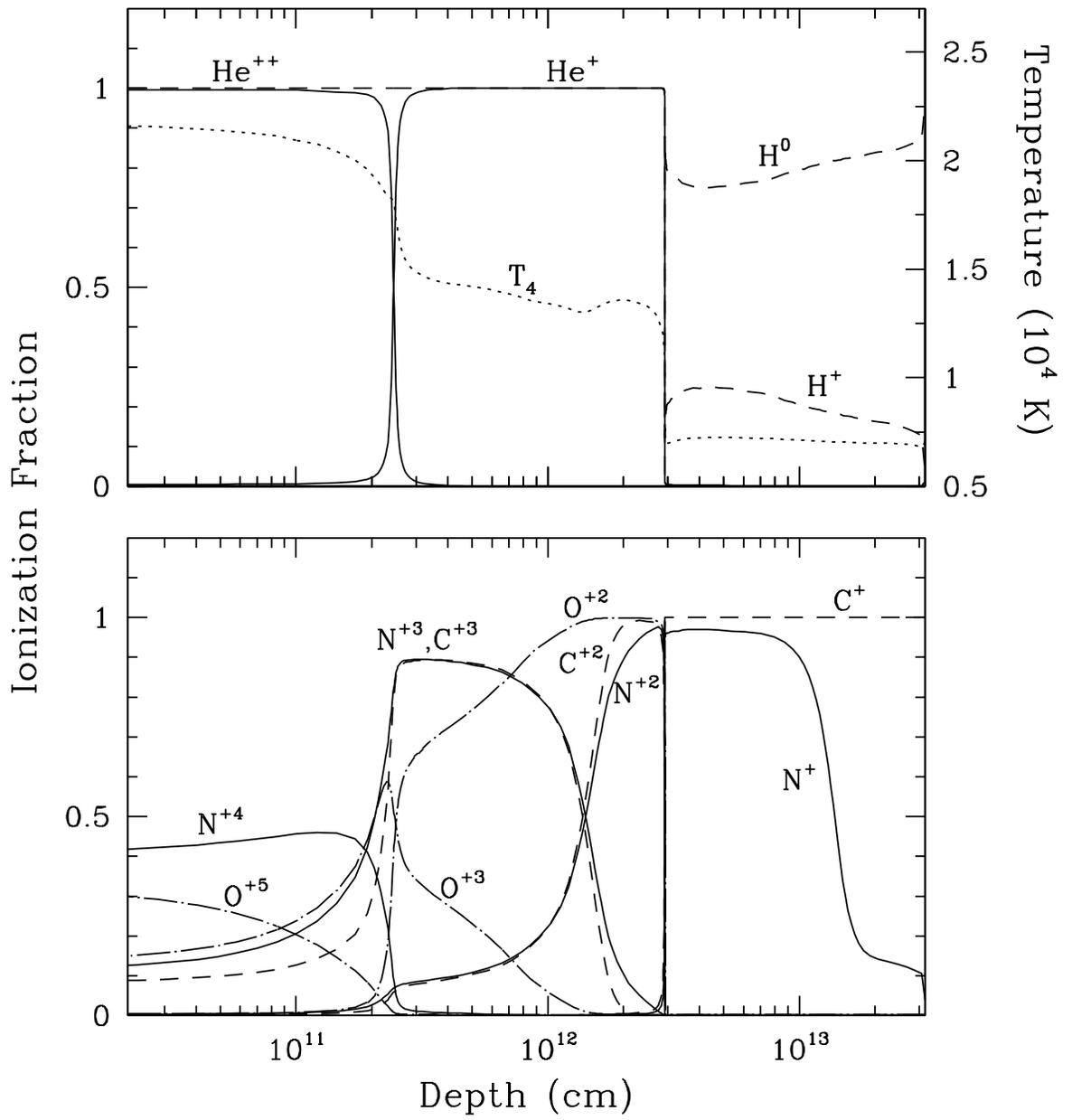
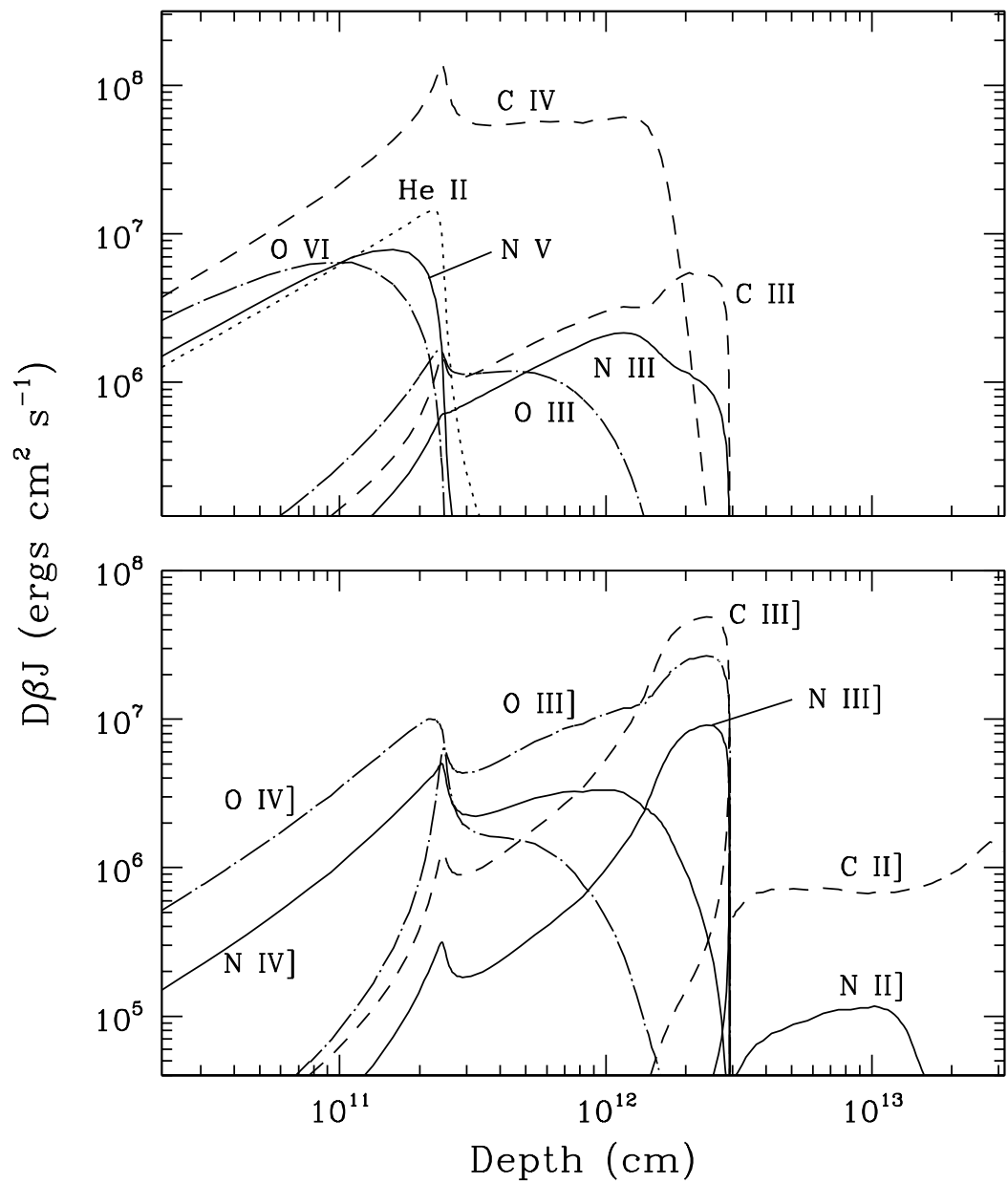
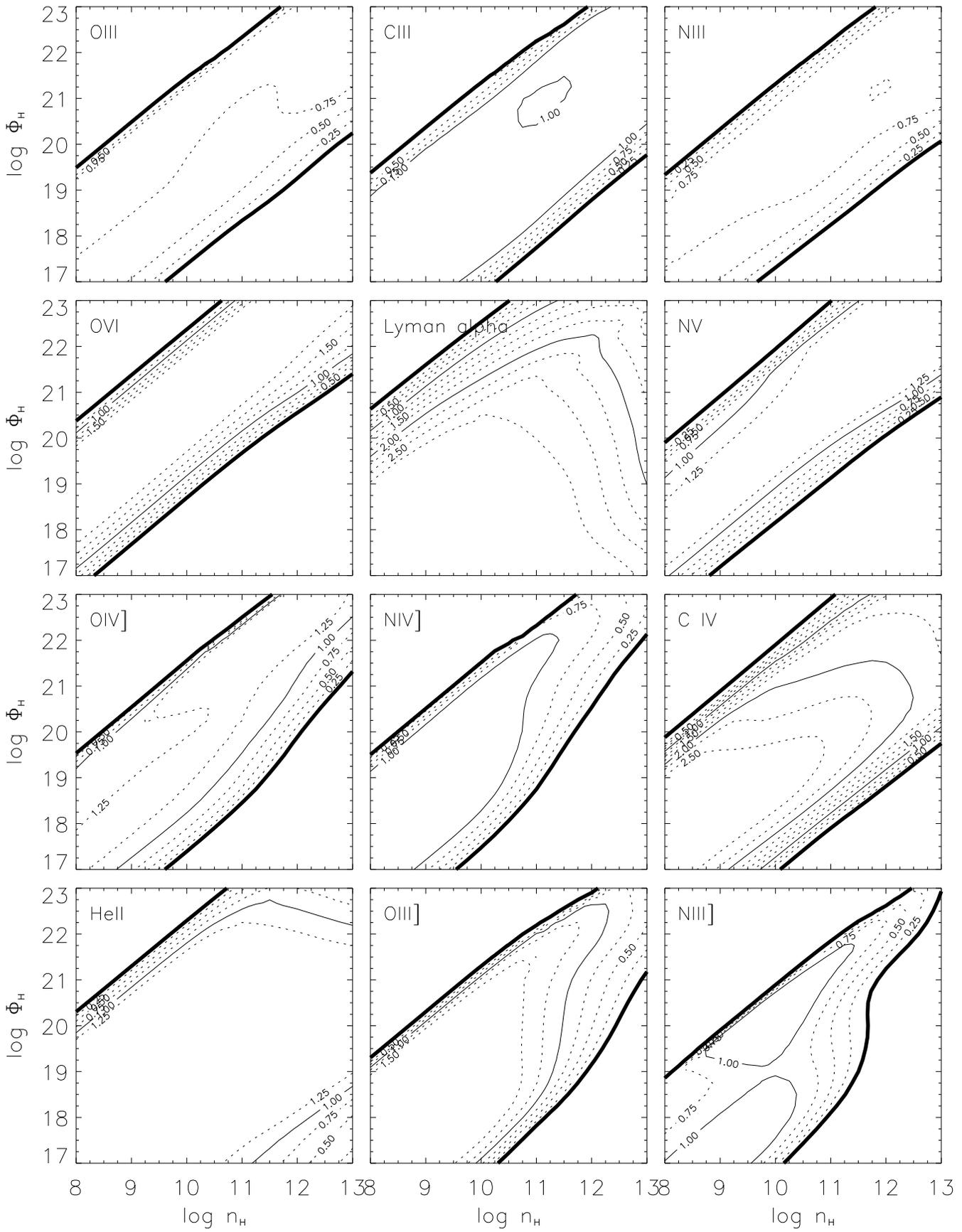
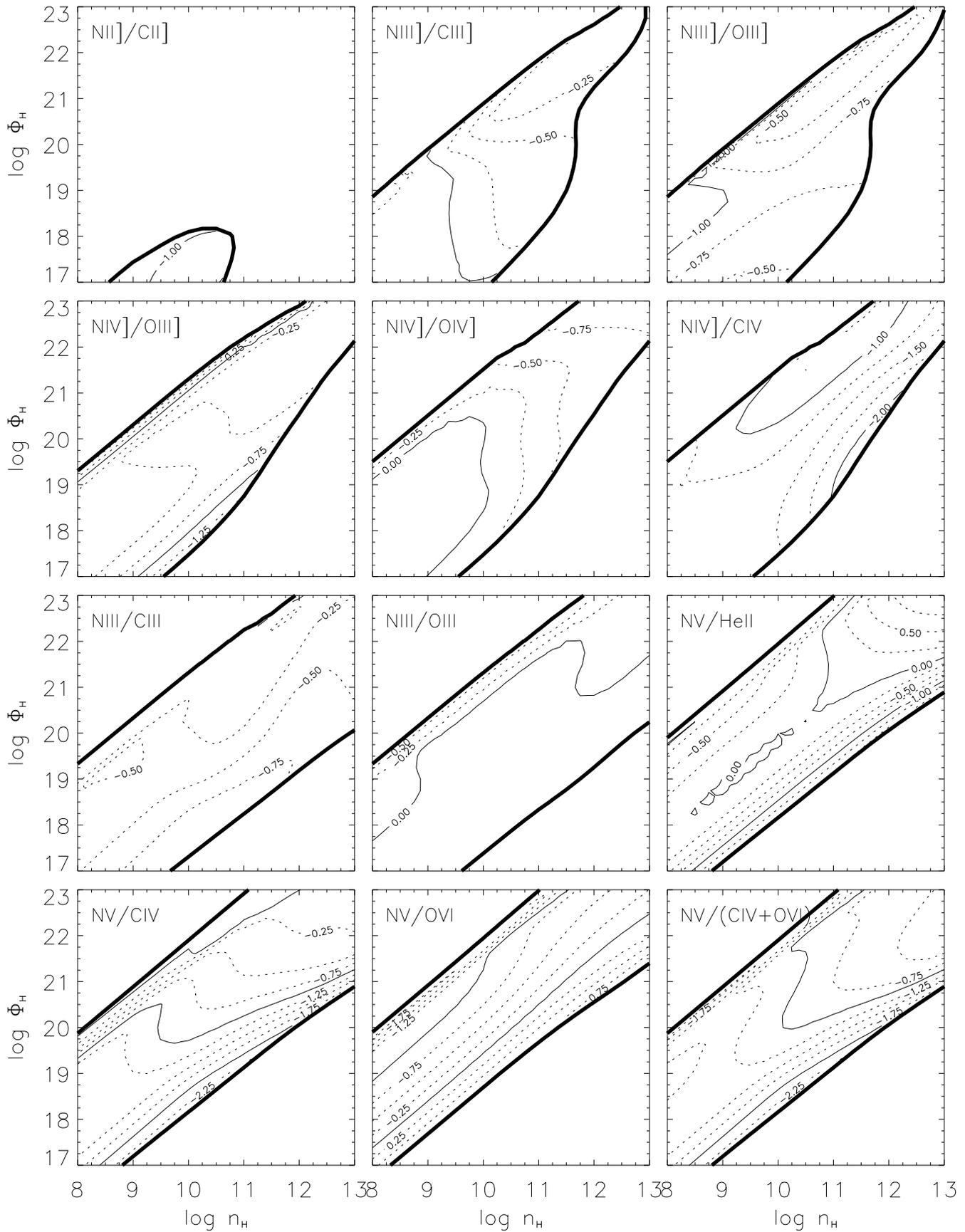
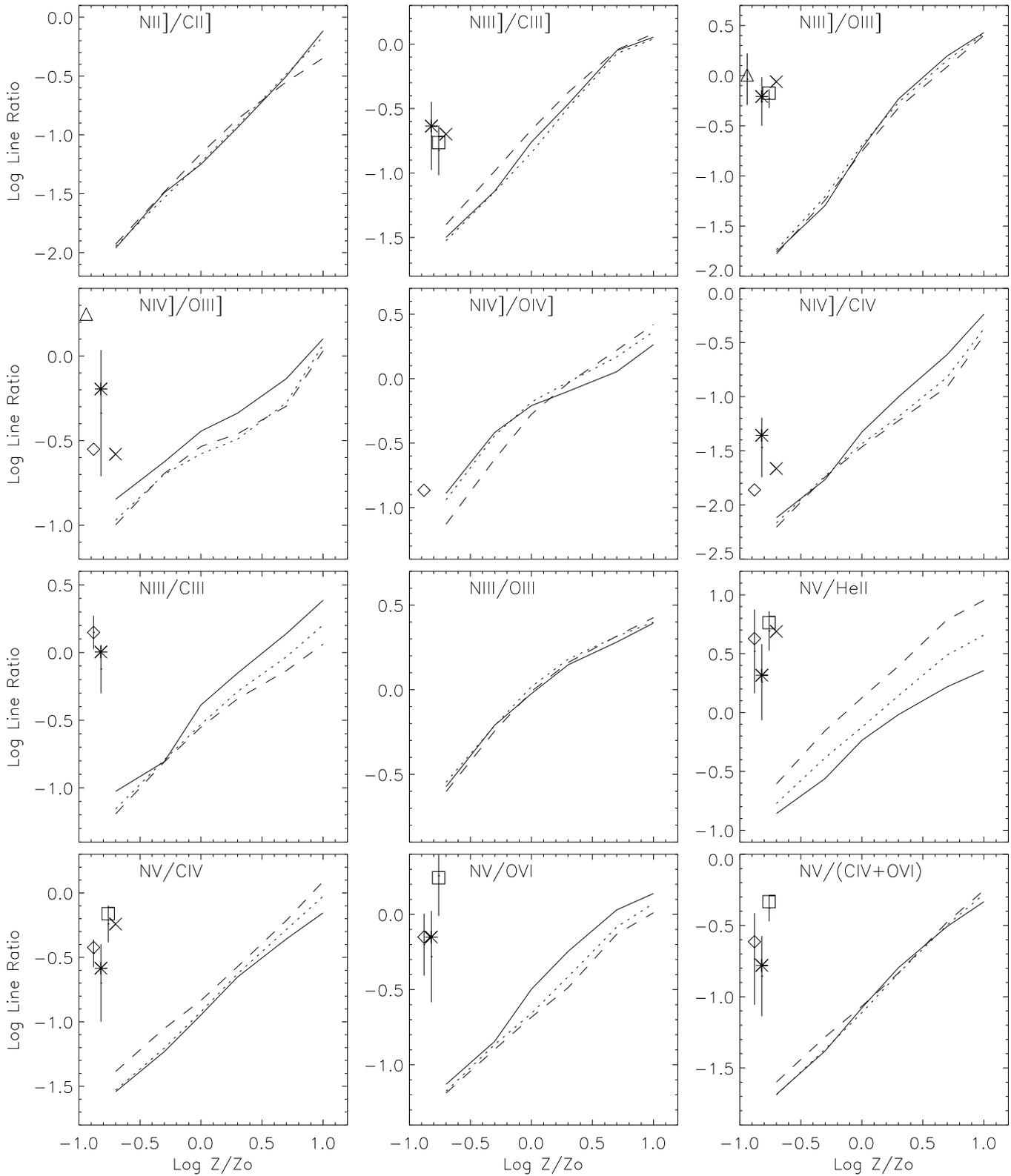
\end{document}